\documentclass{hep99}
\usepackage{times}
\usepackage{epsf}

\begin{document}


\begin{titlepage}

\begin{flushright}
CERN--TH/99-335\\
BUTP--99/21\\ 
MPI-PhT-99-52   \\ 
ZH-TH--27/99\\
hep-ph/99mmnnn                \\[3ex] 
\end{flushright}
\vspace{2.5cm}

\begin{center}
{\large\bf INDIRECT SEARCH FOR SUPERSYMMETRY IN RARE B DECAYS}             \\[5ex] 
 Christoph Greub                                           \\[1ex]
{\it Institut f\"ur theoretische Physik, 
     Universit\"at Bern,   
CH-3012 Bern, Switzerland}       \\[3ex]
 Tobias Hurth                                           \\[1ex]
{\it Theory Division, CERN, CH-1211 Geneva 23, Switzerland\\
    and \\
 Max-Planck-Institut f\"ur Physik, D--80805 M\"unchen, Germany}         \\[3ex]
 Daniel Wyler                                           \\[1ex]
{\it Institut f\"ur theoretische Physik, 
     Universit\"at Z\"urich,
CH-8057 Z\"urich, Switzerland}       \\[7ex]
\end{center}
\begin{center} 
ABSTRACT \\
\vspace*{1mm}
\parbox{13cm}{
QCD corrections to the gluino induced contribution to $b \to s \gamma$
are shown to be important in order to extract reliable bounds on the
off-diagonal elements of the squark mass matrices.}
\end{center}

\vspace*{2truecm}
{\begin{center} 
{\it Talk presented by Tobias Hurth at the International Euroconference} \\
{\it on High Energy Physics (EPS-HEP99), 15-21 July 1999, Tampere, Finland.}
\end{center}}

\vfill
\end{titlepage}

\thispagestyle{empty}


\title{Indirect Search for Supersymmetry\\ in Rare B Decays
}

\author{Christoph Greub$^{\,a}$,
        Tobias Hurth$^{\,b,\,c}$
,\\ and Daniel Wyler$^{\,d}$ }  

%

\address{
$^{a}$ 
 Institut f\"ur Theoretische Physik, Universit\"at Bern, 
 CH--3012 Bern, Switzerland\\
$^{b}$
  Theory Division, CERN, CH--1211 Geneva 23, Switzerland\\
$^{c}$  
 Max-Planck-Institut f\"ur Physik,
 D--80805 M\"unchen, Germany\\ 
$^{d}$
 Institut f\"ur Theoretische Physik, Universit\"at Z\"urich,
 CH--8057 Z\"urich, Switzerland\\}

\abstract{ 
QCD corrections to the gluino induced contribution to $b \to s \gamma$
are shown to be important in order to extract reliable bounds on the
off-diagonal elements of the squark mass matrices.} 

\maketitle
Rare processes are  
an important tool for investigating
new interactions. The standard model contributions are usually small
and new physics may manifest itself clearly. In particular, rare decays  
provide guidelines for supersymmetry model building.   
The experimental observation of these Flavour Changing Neutral
Currents (FCNCs), or the upper limits set on them yield  
stringent relations between the many parameters 
in the soft supersymmetry-breaking terms. The processes
involving transitions between first and second generation quarks,
namely FCNC processes in the $K$ system, are considered to be most 
efficient in shaping viable supersymmetric flavour models. 
Moreover, the tight 
experimental bounds on some flavour diagonal transitions such as the 
electric dipole moment of the electron and of the neutron, as well 
as $g-2$, help constrain soft terms inducing chirality violations. 

The severe experimental constraints on flavour violations have no direct 
explanation in the structure of the minimal supersymmetric
standard model (MSSM). This is the 
essence of the well-known supersymmetric flavour problem.
There exist several supersymmetric models (within the MSSM) with 
specific solutions to this problem. Most popular are the ones  
in which the dynamics of flavour sets in above the supersymmetry 
breaking scale and the flavour problem is killed by the 
mechanisms of communicating supersymmetry breaking to the 
experimentally accessible sector: In the constrained minimal supersymmetric 
standard model (mSUGRA) supergravity is the mediator between the 
supersymmetry breaking and the visible sector \cite{MSUGRA}. 
In  gauge-mediated 
supersymmetry  breaking models (GMSBs) the communication 
between the two sectors is realized by gauge interactions \cite{GMSBs}.
More recently the anomaly mediated 
supersymmetry breaking models (AMSBs) were proposed, in which the  
two sectors  are linked by interactions suppressed by the Planck mass
\cite{ANOMAL}. 
Furthermore, there are other classes of models in which the flavour problem is 
solved by particular flavour symmetries. 

Neutral flavour transitions involving third generation quarks,
typically in the B system, do not pose yet serious threats to these
models. The rare decay $b \rightarrow s \gamma$
has already been detected, but the precision of the measurements 
is at the moment not very high \cite{EXP}. Nevertheless, it has
already carved out some regions in the space of free
parameters of most of the models in the above classes (see \cite{THEO} 
and references therein). In particular, it dangerously constrains  
several somewhat tuned realizations of these
models \cite{TUNED}. Once the experimental precision is 
increased, this decay will undoubtedly help 
selecting the viable regions of the parameter space in the above 
class of models and/or discriminate among these or other possible
models. It is important to calculate the rate of this 
decay with theoretical uncertainties reduced 
as much as possible, and general enough for generic 
supersymmetric models. 
In the standard model, the rate for $b \to s \gamma$
is known
up to  next-to-leading order (NLL) in QCD \cite{NLL}.  
The NLL calculation  reduces the 
large scale dependence present at the LL  ($\pm 25\%$) to a mere 
percent uncertainty. This accuracy is however somewhat fortuitous
as it is obtained through large 
accidential cancellations among different contributions 
to the NLL corrections  \cite{ACC1,ACC2}. 
Indeed, the accuracy for the
NLL calculation of the $b \rightarrow s \gamma$ rate in a two-Higgs
model is substantially worse \cite{ACC1}.

The calculation of this decay rate within supersymmetric models 
is still far from this level of sophistication.
There are several contributions to the decay amplitude: 
Besides the 
$W^- -t$-quark and the $H^- -t$-quark contributions, there are also
the chargino, gluino and neutralino contributions.
All these contributions were calculated in \cite{FRA}
within the mSUGRA model;  
their analytic expressions apply naturally also to the GMSB, and AMSB models. 
The inclusion of QCD corrections needed for the calculation of the rate, 
was in \cite{FRA} assumed to follow the SM pattern. 
A calculation taking into account solely
 the gluino contribution has been performed 
in \cite{MAS} for a generic supersymmetric model, but no QCD
corrections were included. 

An interesting NLL analysis of $b \rightarrow s \gamma$ was recently performed 
\cite{GUD,MIK} in a specific class of 
models where the only source of flavour violation 
at the electroweak scale is that of the SM, 
encoded in the Cabibbo-Kobayashi-Maskawa (CKM) matrix.
This calculation, however,
cannot be used in particular directions of the parameter 
space of the above listed models in which quantum effects induce a 
gluino contribution as large as the chargino or the SM contributions. 
Nor it can be used as a model-discriminator tool, able to constrain 
the potentially
large sources of flavour violation typical of generic 
supersymmetric models.  

Among these, the flavour non-diagonal vertex gluino-quark-squark induced by
the flavour violating scalar mass term and trilinear terms
is particularly interesting. This is generically assumed to induce the 
dominant contribution to quark flavour transitions, as this vertex is weighted
by the strong coupling constant $g_s$.
Therefore, it is often taken as the 
only contribution 
to these transitions and in particular to the $b \rightarrow s \gamma$
decay, when attempting to obtain order-of-magnitude upper bounds
on flavour violating terms in the scalar potential \cite{MAS,HAG}.
Once the constraints coming from the experimental measurements are imposed, 
however, the gluino contribution is reduced to values such that the SM 
and the other supersymmetric contributions cannot be neglected 
anymore. Any LL and NLL calculation of the $b \rightarrow s \gamma$
rate in generic supersymmetric models, therefore, should then include
all possible contributions. 

The gluino contribution, however, presents some peculiar features related
to the implementation of the QCD corrections.
In ref. \cite{OUR} this contribution to the decay $b \rightarrow s \gamma$
is therefore investigated in great detail for
 supersymmetric models with generic soft terms.
It is shown
that 
the relavant operator basis of the SM effective Hamiltonian gets enlarged 
to contain magnetic and chromomagnetic operators with an extra factor of 
$\alpha_s$
 and weighted by a quark mass $m_b$ or $m_c$, and also 
magnetic and chromomagnetic operators of lower dimensionality, as well as 
additional scalar and tensorial 
four-quark operators.
A  few results of our analysis in ref. \cite{OUR} are 
given  in the following, showing the effect of the 
LL QCD corrections on constraints on supersymmetric sources of 
flavour violation.

To understand the sources of flavour violation which may be present in
supersymmetric models in addition to those enclosed in the CKM matrix,
one has to consider the contributions to the squark mass matrices
\begin{equation}
{\cal M}_{f}^2 =  
\left( \begin{array}{cc}
  m^2_{f,LL}   & m^2_{f,LR} \\
  m^2_{f,RL}  &  m^2_{f,RR}                 
 \end{array} \right) +
\label{squarku}
\nonumber
\end{equation}
\begin{equation}
  \left( \begin{array}{cc}
  F_{f,LL} +D_{f,LL} &  F_{f,LR} \\
 F_{f,RL} & F_{f,RR} +D_{f,RR}                
 \end{array} \right) \quad ,
\nonumber
\label{squarku2}
\end{equation}
where $f$ stands for up- or down-type squarks.
In the super  CKM basis where the quark mass matrices are diagonal 
and the squarks are rotated in parallel to their superpartners,
the $F$ terms  from the superpotential and the $D$ terms 
turn out to be diagonal 
$3 \times 3$ submatrices of the 
$6 \times 6$
mass matrices ${\cal M}^2_f$. This is in general not true 
for the additional terms (\ref{squarku}), originating from  the soft 
supersymmetric breaking potential. As a consequence, 
gluino contributions to the
decay $b \to s \gamma$ are induced by the off-diagonal
elements of the soft terms 
$m^2_{f,LL}$, $m^2_{f,RR}$, $m^2_{f,RL}$.

It is convenient to select one possible 
source of flavour violation in the squark sector at a time and
assume that all the remaining ones are vanishing. Following
ref.~\cite{MAS}, all diagonal entries in 
$m^2_{\,d,\,LL}$, $m^2_{\,d,\,RR}$, and $m^2_{\,u,\,RR}$
are set to be equal and their common value is denoted by
$m_{\tilde{q}}^2$.  The branching ratio can then be studied as a
function of 
\begin{equation} 
\delta_{LL,ij} = \frac{(m^2_{\,d,\,LL})_{ij}}{m^2_{\tilde{q}}}\,, 
\hspace{0.1truecm}
\delta_{RR,ij} = \frac{(m^2_{\,d,\,RR})_{ij}}{m^2_{\tilde{q}}}\,, 
\hspace{0.1truecm} 
\label{deltadefa}
\end{equation}
\begin{equation} 
\delta_{LR,ij} = \frac{(m^2_{\,d,\,LR})_{ij}}{m^2_{\tilde{q}}}\,, 
\, (i \ne j).
\label{deltadefb}
\end{equation}
The remaining crucial parameter needed to determine the 
branching ratio is $x = m^2_{\tilde{g}}/m^2_{\tilde{q}}$,
where $m_{\tilde{g}}$ is the gluino mass.
In the following, we concentrate on the LL QCD corrections to the 
gluino contribution. 
\begin{figure}[ht]
\begin{center}
\leavevmode
\epsfxsize= 7.5 truecm
\epsfbox[18 167 580 580]{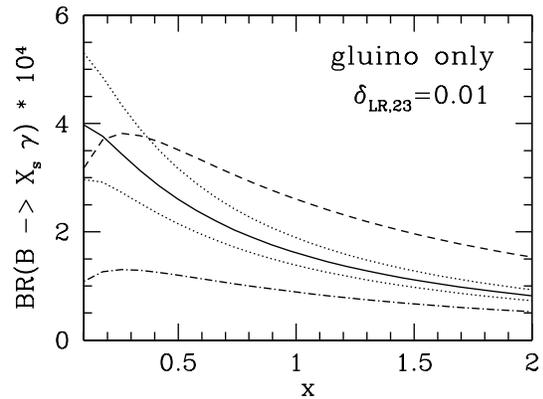}
\end{center}
\caption[f1]{Gluino-induced branching ratio $BR(B \to X_s \gamma)$ 
 as a function of $x= m^2_{\tilde{g}}/m^2_{\tilde{q}}$, obtained when
 the only source of flavour violation is $\delta_{LR,23}$ (see text).
}
\label{sizeqcd23lr}
 \end{figure}
 \begin{figure}[th]
\begin{center}
\leavevmode
\epsfxsize= 7.5 truecm
\epsfbox[18 167 580 580]{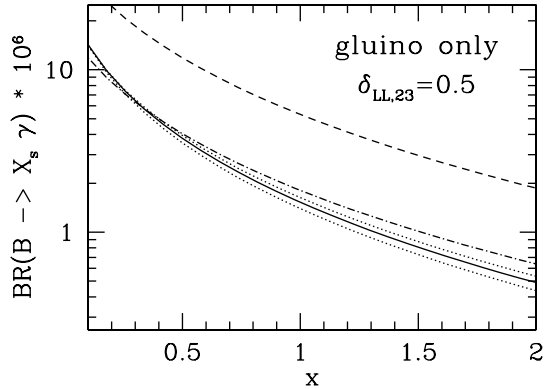}
\end{center}
\caption[f1]{Same as in Fig.~\ref{sizeqcd23lr} when only
 $\delta_{LL,23}$ is non-vanishing.}
\label{sizeqcd23ll}
\end{figure}
In Figs.~\ref{sizeqcd23lr} and~\ref{sizeqcd23ll},
the solid lines show  the QCD corrected
branching ratio, when only $\delta_{LR,23}$ or
$\delta_{LL,23}$ are non vanishing.  
The branching ratio is plotted as a function of
$x$, using
$m_{\tilde{q}}=500\,$GeV.  The  dotted lines show the range of variation
of the branching ratio, when the
renormalization scale $\mu$ varies in the interval $2.4$--$9.6\,$GeV. 
Numerically, the scale uncertaintly in
$BR(B \to X_s \gamma)$ is about
$\pm 25\%$. An extraction of bounds on the $\delta$
quantities more precise than just an order of magnitude, therefore,
would require the inclusion of next-to-leading logarithmic QCD
corrections. It should be noticed, however, that the inclusion of the
LL QCD corrections has already removed the large ambiguity on the
value to be assigned to the factor $\alpha_s(\mu)$ in the
gluino-induced operators. Before
adding QCD corrections, the scale in this factor can assume all values
from $O(m_b)$ to $O(m_W)$: the difference between $BR(B \to X_s \gamma)$
obtained when $\alpha_s(m_b)$ or when $\alpha_s(m_W)$ is used, is
of the same order as the LL QCD corrections.  The corresponding values
for $BR(B \to X_s \gamma)$ for the two extreme choices of $\mu$ are
indicated in Figs.~\ref{sizeqcd23lr} and~\ref{sizeqcd23ll} by
the dot-dashed lines ($\mu=m_W$) and the dashed lines
($\mu=m_b$). The
choice $\mu = m_W$ gives values for the non-QCD corrected 
$BR(B \to X_s \gamma)$ relatively close to the band obtained when the
LL QCD corrections are included, if only $\delta_{LL,23}$ is
non-vanishing. Finding a corresponding value of $\mu$ that minimizes
the QCD corrections in the case studied in Fig.~\ref{sizeqcd23lr},
when only $\delta_{LR,23}$ is different from zero,
depends strongly on the value of $x$.
In the context of the full LL result, it is important
to stress that the explicit $\alpha_s$ factor 
                       has to be evaluated 
- like the Wilson coefficients - at a scale $\mu=O(m_b)$.

In spite of the large uncertainties which the branching ratio 
$BR(B \to X_s \gamma)$ still has at  LL in QCD, it is possible
to extract indications on the size that the $\delta$-quantities 
may maximally acquire without inducing conflicts with the 
experimental measurements (see \cite{OUR}). 

\vspace{0.7cm}
\noindent
\underline{Acknowledgment:}
This work was partially supported by Schweizerischer Nationalfonds.

\end{document}